# Ultrastrong and Ultrastable Metallic Glass

Daisman P.B. Aji,[1] Akihiko Hirata,[1] Fan Zhu,[1] Liu Pan,[1] K. Madhav Reddy,[1] Shuangxi Song,[2] Yanhui Liu,[1] Takeshi Fujita,[1] Shinji Kohara[3] and Mingwei Chen[1,2,]*

[1] WPI-Advanced Institute for Materials Research (WPI-AIMR), Tohoku University

Sendai 980-8577, Japan

[2] State Key Laboratory of Metal Matrix Composites, School of Materials Science and

Engineering, Shanghai Jiao Tong University, Shanghai 200030, PR China

[3] Japan Synchrotron Radiation Facility/SPring-8, Hyogo 679-5198, Japan

*The lack of thermal stability, originating from their metastable nature, has been one of the paramount obstacles that hinder the wide range of applications of metallic glasses. We report that the stability of a metallic glass can be dramatically improved by slow deposition at high temperatures. The glass transition and crystallization temperatures of the ultrastable metallic glass can be increased by 51 K and 203 K, respectively, from its ordinary glass state. The ultrastable metallic glass also shows ultrahigh strength and hardness, over 30 % higher than its ordinary counterpart. Atomic structure characterization reveals that the exceptional properties of the ultrastable glass are associated with abundance of medium range order. The finding of the ultrastable metallic glass sheds light on atomic mechanisms of metallic glass formation and has important impact on the technological applications of metallic glasses.*

Multicomponent bulk metallic glasses are among the most unique and fascinating of the recently discovered materials. *(1-6)* However, their many technological applications are hindered by low thermal stability because of the metastable nature toward crystallization at high temperatures or during long-time services at low temperature. The stability of metallic glasses is commonly expressed with a measure, namely glass forming ability (GFA); the higher is the GFA, the more stable is the glass. GFA has a linear tendency in two empirical parameters *(3, 7, 8)*: *1)* reduced glass transition temperature, $T_{rg}$, defined as $T_g/T_m$, where $T_g$ is the glass transition temperature and $T_m$ is the melting temperature, and *2)* width of supercooled liquid region, $\Delta T_x$, defined as $\Delta T_x = T_x - T_g$, where $T_x$ is the crystallization temperature. GFA may also be interchangeably called with a term 'thermal stability', which is a measure of how easy a glass to crystallize upon heating from glassy state to supercooled liquid state or upon isothermal annealing. Compared to organic and oxide glasses, metallic glasses usually have much low stability and GFA because of high atomic mobility and resultant fast structure relaxation at high temperatures. The lack of thermal stability has been one of the major obstacles that hinder the wide range of applications of metallic glasses.

A new paradigm of glass stability in terms of kinetics and thermodynamics has recently emerged with the discovery of ultrastable organic glasses by Ediger and his co-workers. *(9)* The $T_g$ of the stable organic glasses increases obviously for 16 K compared to that of the ordinary glasses. The stable glasses also have much lower enthalpy and fictive temperature, $T_f$, than the ordinary and annealed ordinary glasses. The key point to obtain such high stability is the very slow deposition rates at the optimum high temperatures to allow arriving molecules on the substrate to have sufficient time to rearrange themselves for high efficient packing. Despite the fact that many ultrastable organic glasses have been discovered *(9-15)* and stable Lennard-Jones glasses has also been predicted by recent molecular dynamics (MD) simulation *(16)*, ultrastable metallic glasses have not been experimentally realized so far.

In this study we report that an ultrastable metallic glass can be achieved from a multicomponent $Zr_{55}Cu_{30}Ni_5Al_{10}$ (at %) alloy using a single-target magnetron sputtering method. *(17, 18)* By tuning the sputtering rate and deposition temperature, we are able to obtain an exceptionally stable metallic glass with an increase in $T_g$ for 51 K (from 716 K of ordinary glass to 767 K of the ultrastable glass) and $T_x$ for 207 K (from 753 K of ordinary to 960 K of ultrastable glass). The improved kinetic and thermodynamic stability is much more significant than that obtained from ultrastable organic metallic glasses that only have the maximum $T_g$ increment of ~40 K. *(14)* The ultrastable metallic glass also has improved mechanical properties. The hardness and fracture strength are ~10.46 GPa and ~3.0 GPa, respectively, which are ~30 % higher than the ordinary metallic glass. With the discovery of this ultrastable and ultrastrong metallic glass, the glass that can be obtained with high kinetic and thermodynamic stability is not exclusive only to organic glasses and the deposition technique may be used as a general approach to produce ultrastable glasses in any type of amorphous materials.

**Figure 1A** shows the differential scanning calorimetry (DSC) profiles at a heating rate of 20 K/min for bulk and two films of $Zr_{55}Cu_{30}Ni_5Al_{10}$ glass samples. The bulk glass was produced by copper-mold-casting. The two film glasses with a thickness of ~30 μm were produced by rf magnetron sputtering at deposition temperatures ($T_{dep}$) and rates ($V_{dep}$) of 300 K and 0.25 nm/s as well as 573 K and 0.19 nm/s, respectively. As shown in **Table 1** and **Fig. 1A**, the glass transition temperatures of the bulk and $T_{dep}$ =300 K film glasses are almost the same at ~716 K while the $T_g$ of the film with $T_{dep}$ at 573 K is 767

K, a remarkable increase of ~51 K. In addition to the high $T_g$, the $T_{dep}$ = 573 K glass also shows high stability against crystallization with $T_x$ of 960 K, about 207 K higher than those of bulk and $T_{dep}$ =300 K film glass, and a enhanced supercooled liquid region $\Delta T_x$ of 193 K, which is about 4-5 times larger than those of the bulk and $T_{dep}$ =300 K film glasses. The increase in $T_g$, $T_x$ and $\Delta T_x$ of the high-temperature deposited film demonstrates high kinetic stability of this glass; henceforth this glass is referred to ultrastable glass while the film with $T_{dep}$ of 300 K is referred to ordinary glass. Importantly, the liquidus temperatures of the three glass samples are nearly identical (**Table 1** and **Fig. 1A**), validating that the extraordinary stability of the ultrastable glass is not associated with chemical composition variation.

**Figure 1B** shows the X-ray diffraction (XRD) profiles of the bulk, ordinary and ultrastable glasses. The diffraction angle of the first peak of the ultrastable metallic glass is slightly smaller than those of the bulk and ordinary glasses, as shown in **Table 1**. The sharp diffraction peaks are from the Al substrate, which also act as the internal marks to characterize the peak positions of the glassy samples. The peak shift to lower diffraction angles was also observed in stable organic glasses. *(12)* Since the compositions of the bulk, ordinary and ultrastable glass are essentially the same, verified by energy dispersive spectroscopy (EDS) and the DSC liquidus temperature, the peak shift can only be interpreted by the difference in atomic configuration between the ultrastable glass and the bulk and ordinary ones. The full width at the half maximum (FWHM) of the peak of the ultrastable glass is broader than that of the bulk and ordinary glasses, similar with the stable organic glasses too *(12),* indicating a wide spread of the interatomic distance in the ultrastable glass. We also utilized synchrotron radiation XRD to investigate the atomic structure of the ultrastable and ordinary glasses. **Fig. 1C** displays the pair distribution functions (PDF), $g(r)$, derived from Fourier transform of structural factors. Close inspection indicates that the first peaks of the $Zr_{55}Cu_{30}Ni_5Al_{10}$ glasses split into two sub-peaks locating at ~2.80 Å and 3.15 Å (see the inset of **Fig. 1C**) due to the discrete bonding length distribution of the nearest neighbor atoms. According to the Goldschmidt radii and weight factors (**Table S1**), the first sub-peak can be assigned mainly to the Zr-Cu pair, while the Zr-Zr pair dominates the second sub-peak. By integrating the first peak of the PDFs from 2 Å to 3.9 Å, the average coordination numbers are found to be constant with a value of ~13.6 for the two glasses, while the ultrastable glass has a higher weight fraction of the Zr-Cu pair and lower one of Zr-Zr pair compared to the ordinary glass, suggesting that chemical ordering takes place in the ultrastable glass. Moreover, the peak of the Zr-Zr pair slightly shifts to a larger bonding distance (see the inset of **Fig. 1C**). Therefore, the broadening and low-angle shift of the XRD peak of the ultrastable metallic glass, shown in **Fig. 1B,** is apparently associated with the chemical ordering with the formation of Zr-Cu pair enriched domains with a short interatomic bonding length and Zr-Zr pair rich domains with a relatively longer bonding length.

**Figure 2A** shows DSC scans at heating rate of 20 K/min of bulk and ultrastable $Zr_{55}Cu_{30}Ni_5Al_{10}$ glass. The bulk glass was annealed at 683 K (0.95 $T_g$) for 60 (red) and 360 min (blue). The heating curves of the annealed samples show endothermic peak due to the regain of enthalpy loss during annealing, however the glass transition temperatures of the annealed samples remain unchanged compared to the unannealed one, suggesting that the structural relaxation by annealing cannot obviously enhance the kinetic stability of metallic glasses. **Fig. 2B** shows the enthalpy curves of the $Zr_{55}Cu_{30}Ni_5Al_{10}$ glass samples derived by integrating the curves shown in **Fig. 2A**. It is obvious that the enthalpy of ultrastable glass is much lower than the bulk and annealed bulk glasses. The low energy state of the ultrastable glass can also be seen from the low fictive temperature, $T_f$, of 678 K, which is ~38 K lower than that of the bulk metallic glass. Interestingly, the ultrastable glass can transform into an ordinary glass after annealing at 893 K (in the supercooled liquid region of the ultrastable glass) for 2 min then cooled at 20 K/min to room temperature; thereafter the sample was heated at 20 K/min rate. The heating curve was shown in **Fig. 2C** (blue curve). The new $T_g$ (blue curve) of the annealed ultrastable glass is ~711 K, close to the $T_g$ of the bulk glass (716 K). The transformation from ultrastable glass to the ordinary one implies that the exceptional stability of the ultrastable metallic glasses derives from the unique atomic structure formed by high-temperature slow deposition. The thermal stability of the ultrastable metallic glass was also tested by annealing experiments. Two samples of the ultrastable glass were annealed at 683 K (0.95$T_g$) for 60 and 360 min. The heating curves of the annealed samples (**Fig. 2D**) do not show endothermic peaks expected from the enthalpy regain due to annealing.

The mechanical properties of the ultrastable glass were characterized by nanoindentation and microcompression tests. The nanoindentation force-depth curves (**Fig. 3A**) suggest that the ultrastable glass is much stronger against deformation than bulk and ordinary ones. As shown in **Fig. 3B,** the measured hardness and elastic modulus of the ultrastable glass are ~ 10.46 ± 0.58 GPa and 156.75 ± 7.54 GPa, respectively, which are over 30% higher than those of the bulk (Hardness: 6.69 ± 0.24 GPa. Elastic modulus: 106.44 ± 2.91 GPa) and ordinary glasses (Hardness: 6.99 ± 0.12 GPa. Elastic modulus: 109.76 ± 1.94 GPa). It is interesting to note that the increase in elastic modulus was also observed in a stable organic glass. *(19)* The ultrahigh strength of the ultrastable glass was further confirmed by uniaxial microcompression tests (**Fig. 3C** and **D**). The failure strength of the ultrastable glass is ~3.0 GPa, ~1.0 GPa higher than that of the ordinary glass. The deformation behavior of the ultrastable glass is similar to that of ordinary metallic glasses and obvious shear bands can be seen from the tested micropillars although shear bands are difficult to find in the vicinity of the residual impressions of nanoindentation.

**Figure 4A** is the high-resolution transmission electron microscope (HRTEM) image of the ultrastable glass. Obvious periodic contrast cannot be seen form the phase-contrast micrograph, demonstrating that the ultrastable glass has a disordered amorphous structure, similar to ordinary metallic glasses. The corresponding selected area electron diffraction (SAED) pattern (**Fig. 4B**), taken from a large selected area, further demonstrates the amorphous nature of the ultrastable glass. The main diffraction peaks in the SAED intensity profile are in good agreement with those of the XRD spectrum (**Fig. 1B**), indicating the structure of the TEM foil is the same as those of thick ultrastable glass samples that were used in DSC and XRD measurements. Different from the phase-contrast HRTEM image, the scanning transmission electron microscopy (STEM) is more susceptible to local structure disparity and can project the local atomic arrangements in real space. The bright-field and high-angle annular dark field (HAADF) STEM micrographs (**Fig. 4C** and **D**) show the obvious contrast variation with ordered domains of ~1-2 nm in the ultrastable glass. Since the contrast of the HAADF image is associated with the atomic numbers of constitute elements, the ordered domains apparently correlate with the chemical ordering as revealed by the PDF profile (**Fig. 1C**). However, regardless of the bright (presumably Zr-rich) and dark (Cu-rich) regions, highly ordered local structure can be identified from the real-space image, suggesting pronounced medium-range order (MRO) in the ultrastable metallic glass. Angstrom beam electron diffraction (ABED) was employed to characterize the atomic structure of the chemically ordered domains. *(20)* Quantitative measurements of the diffraction vectors of the ABED patterns taken from the ordered domains demonstrate that the local ordering is not consistent with any known simple crystal structure that may appear in the alloy. Instead, most diffraction patterns are compatible with those of the prevailing Voronoi polyhedra in the quaternary alloy predicted by MD simulations. Particularly, the MRO domains often show the structure of <0 2 8 1> or <0 0 12 0>, which are distorted icosahedra with local crystal-like cubic symmetry (**Fig. S1**). *(21)* Although the distorted icosahedral clusters have been frequently observed in ordinary metallic glasses, they usually appear as short-range order. The formation of the distorted icosahedral MRO is most likely the structural origin of the ultrastable metallic glass.

MRO in metallic glasses is usually viewed as a group of atomic clusters that are tightly packed with crystal order or icosahedral order *(22, 23)*. The distorted icosahedral MRO with local cubic symmetry observed in the ultrastable metallic glasses provides new insights on the intrinsic correlation between MRO and glass stability, in addition to previous theoretical and experimental accomplishments. *(22-26)* Since MRO usually includes tens to hundreds of cooperative atoms that mostly sit at the energy minima, it is impractical to form a large number of the distorted icosahedral MRO by slowly cooling a liquid or annealing a glass at high temperatures. On the other hand, high-temperature slow deposition gives deposited atoms sufficient time to rearrange themselves on sample surfaces for high efficient MRO packing before they are buried by later arriving atoms. The low-energy MRO with denser atomic packing is expected to give rise to the high stability of the ultrastable glasses because higher kinetic energy is required to activate or change the MRO into a supercooled liquid state. The dense atomic packing has been confirmed by macroscopic density measurements. There is ~0.5% difference in the density between the ordinary (6.89 g/cm$^3$) and ultrastable glasses (6.93 g/cm$^3$), measured by an X-ray reflectivity method. The higher density from dense atomic packing also offers a straightforward explanation on the significant increase in the mechanical and thermal stabilities of the ultrastable glass.

We have utilized the same technique to prepare binary and ternary ultrastable metallic glasses. However, the improvement in $T_g$ and $T_x$ is not as significant as the multicomponent alloy and crystalline phases often appear during high-temperature deposition. Different from organic and molecular glasses, metal atoms usually have high tendency to form low-energy crystals. Therefore, multiple components of precursor alloys appear to be critical for the formation of ultrastable metallic glasses. This may be because the highly mixed atoms cannot find enough partners to grow as crystals but only form the MRO structure in the sputtering time scale at high temperatures.

In summary, we have developed a $Zr_{55}Cu_{30}Ni_5Al_{10}$ glass with exceptional kinetic and thermodynamic stability by using single-target rf magnetron sputtering deposition at a high temperature and slow deposition rate. The discovery of the ultrastable metallic glass has important implications in understanding glass and glass phenomena and will also be noteworthy for promoting the technological applications of metallic glasses in MEMS, surface coating, electronic and magnetic devices and so on, where thermal and mechanical stability is critical.

**Acknowledgements**

We thank Prof. John Perepezko and Prof. Takeshi Egami for stimulating discussions. This work was sponsored by ''World Premier International Research Center (WPI) initiatives for Atoms, Molecules and Materials,'' MEXT, Japan, and National Natural Science Foundation of China (NSFC, 51271113). We also thank the Center for Computational Materials Science at the Institute for Materials Research, Tohoku University, and SPring-8, Japanese Synchrotron Radiation Facility, Hyogo, Japan under proposal number 2013A1539.

Correspondence and requests for materials should be addressed to M.W.C (mwchen@wpi-aimr.tohoku.ac.jp).

**Table 1.** Glass transition temperature ($T_g$), crystallization temperature ($T_x$), liquidus temperature ($T_l$), supercooled liquid interval ($\Delta T_x$), nominal cooling rate ($Q$), 1st- and 2nd-peak and full width at the half maximum (FWHM) of XRD profiles, hardness, and reduced elastic modulus ($E$) of $Zr_{55}Cu_{30}Ni_5Al_{10}$ glass.

| Sample | $T_g$ (K) | $T_x$ (K) | $T_l$ (K) | $\Delta T_x$ (K) | $Q$ (K/s) | 1st Peak (deg.) | 1st Peak FWHM (deg.) | 2nd Peak (deg.) | 2nd Peak FWHM (deg.) | Hardness (GPa) | $E$ (GPa) |
|---|---|---|---|---|---|---|---|---|---|---|---|
| Bulk | 716 | 753 | 1129 | 37 | 5.8 x 10³ | 37.56 | 5.86 | 64.38 | 11.84 | 6.69 ± 0.24 | 106.44 ± 2.91 |
| $T_{dep}$ = 300 K (ordinary) | 716 | 757 | 1132 | 41 | 9.8 x 10⁸ | 37.51 | 6.55 | 64.45 | 11.79 | 6.99 ± 0.12 | 109.76 ± 1.94 |
| $T_{dep}$ = 573 K (ultrastable) | 767 | 960 | 1142 | 193 | 4.7 x 10³ | 36.54 | 9.11 | 65.84 | 9.50 | 10.46 ± 0.58 | 156.75 ± 7.54 |

**Caption for Figures**:

**Fig. 1.** DSC scans at heating rate of 20 K/min of $Zr_{55}Cu_{30}Ni_5Al_{10}$ glass samples: **(A),** bulk glass produced by copper-mold-casting (black); ordinary and ultrastable glass produced by RF sputtering deposited at 300 (red) and 573 K (green), respectively. The deposition rates for ordinary and ultrastable glass are 0.25 and 0.19 nm/s, respectively. **(B)** The corresponding x-ray diffraction patterns. The four sharp peaks observed on the patterns originate from aluminum pans as substrate. **(C)** Pair distribution functions for ordinary and ultrastable glass. Inset shows the first peaks.

**Fig. 2.** DSC scans at heating rate of 20 K/min of $Zr_{55}Cu_{30}Ni_5Al_{10}$ glass samples: **(A)** bulk glass produced by copper-mold-casting (black); ultrastable glass produced by RF sputtering deposited at 573 K (green); bulk glass annealed at 683 K for 60 (red) and 360 min (blue). **(B)** Enthalpy of $Zr_{55}Cu_{30}Ni_5Al_{10}$ glass samples derived by integrating the curves shown in **(A)**. **(C)** bulk glass (red); as-prepared ultrastable glass (green); ultrastable glass heated to supercooled liquid (893 K) and cooled at 20 K/min rate to room temperature (blue). **(D)** ultrastable glass (green) and ultrastable glass annealed at 683 K for 60 (red) and 360 min (blue).

**Fig. 3.** **(A)** Force-depth curves of $Zr_{55}Cu_{30}Ni_5Al_{10}$ glass samples: bulk glass (black); ordinary and ultrastable glasses deposited at 300 (red) and 573 K (green), respectively. **(B)** The corresponding hardness and reduced elastic modulus, as shown in Table 1, determined from the force-depth curves. **(C)** Engineering stress *vs* strain of ordinary and ultrastable glass. **(D)** Image of micropillars of ultrastable glass after compression.

**Fig. 4.** Electron micrographs and diffraction profiles of ultrastable metallic glass. **(A)** HREM image. No obvious periodic contrast can be observed, demonstrating that the sample has a disorder structure down to a sub-nanometer scale. **(B)** The electron diffraction (blue curve and inset) and synchrotron XRD (red curve) profile. S(Q) is the structure factor and Q is the scattering vector; $Q = 4\pi \sin(\theta)/\lambda$. The * marked peak in the intensity profile is from the amorphous carbon substrate. **(C)** BF-STEM image. **(D)** HAADF-STEM image. The BF- and HAADF- images show that the structure of the glass has a number of MROs with size of ~2 nm.

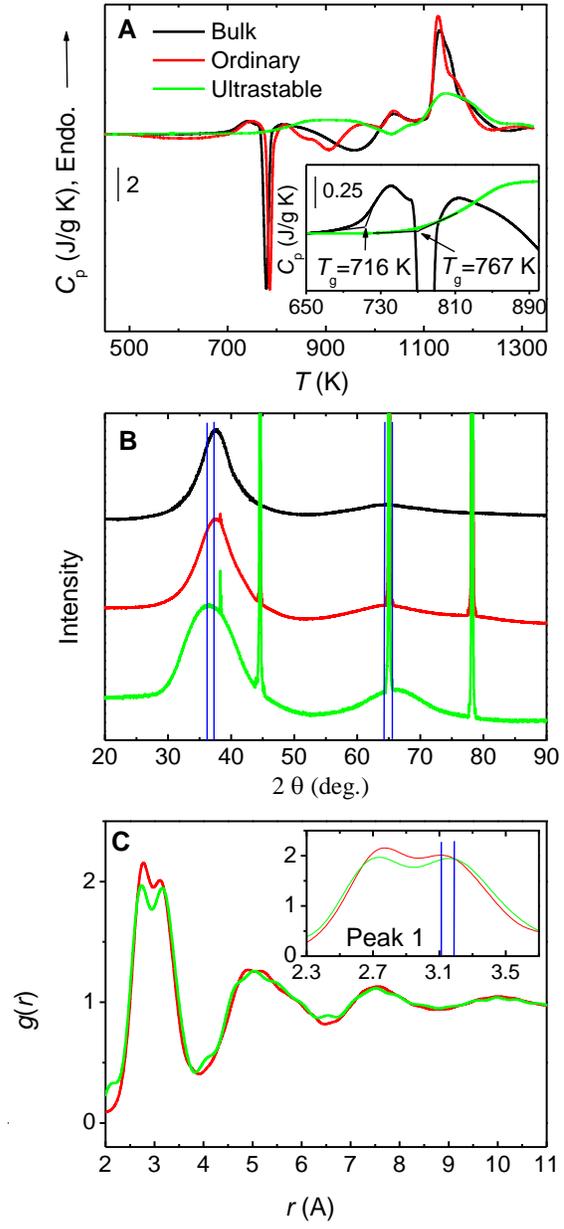

**Fig. 1. Aji et. al.**

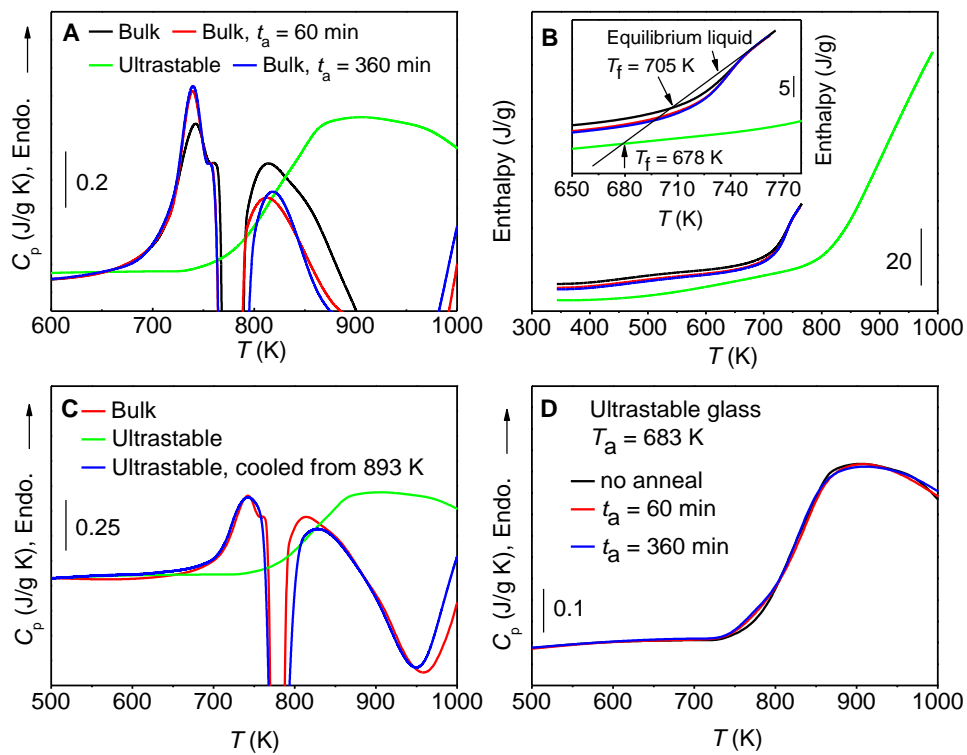

**Fig. 2. Aji et. al.**

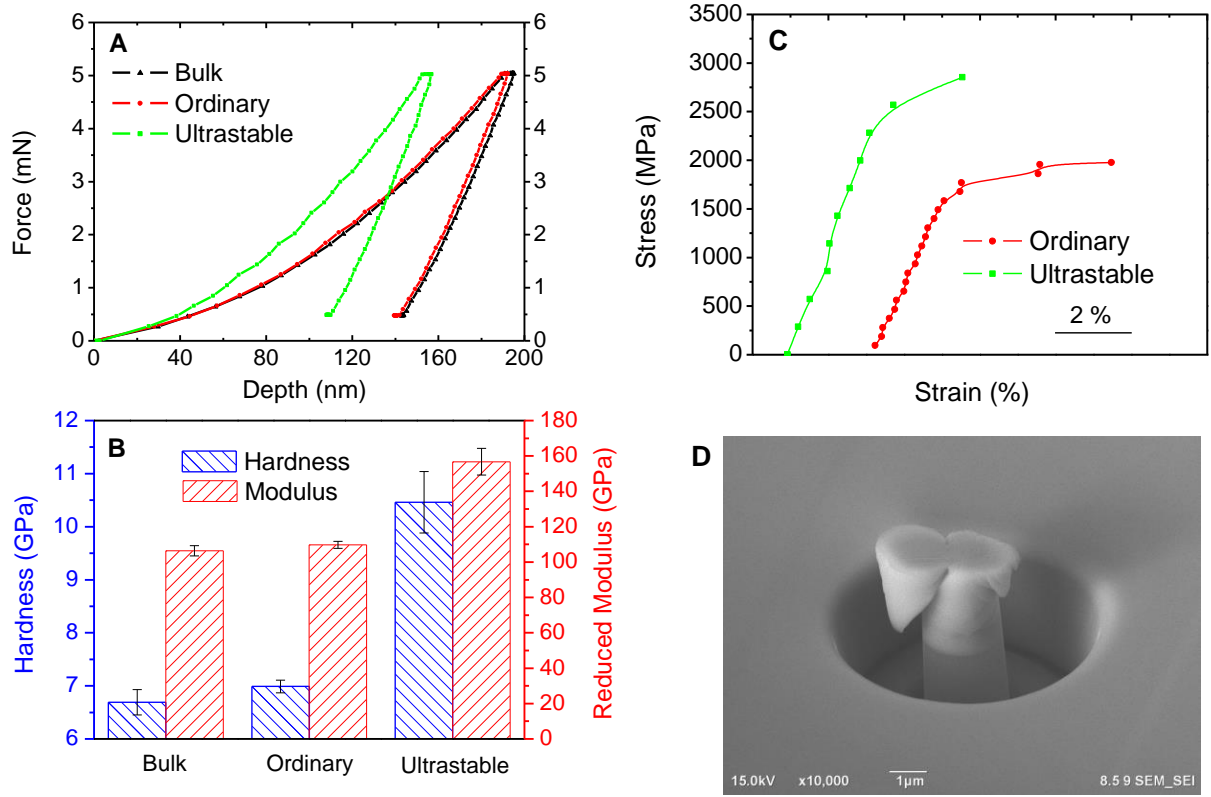

**Fig. 3. Aji et. al.**

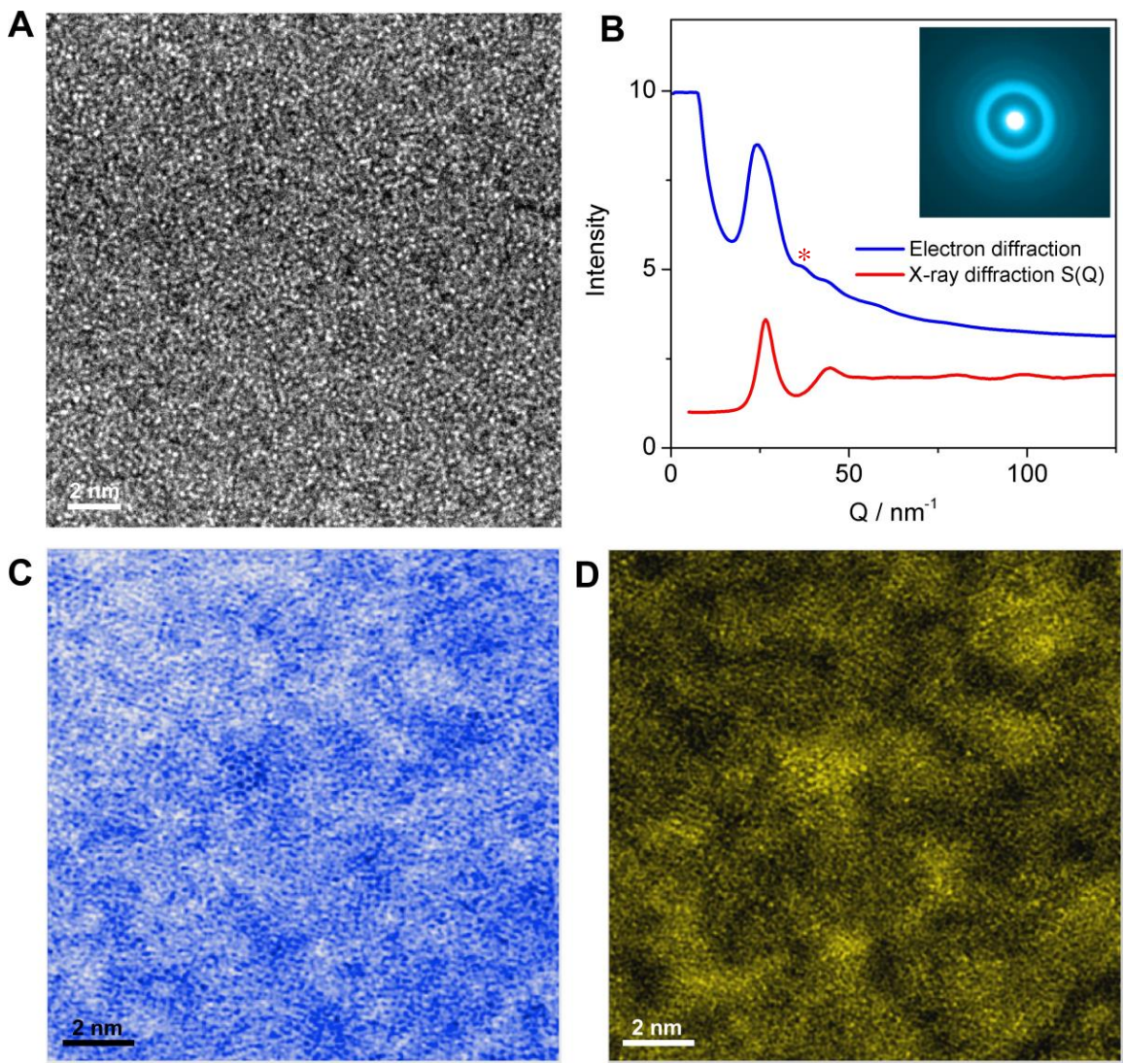

**Fig. 4. Aji et. al.**

# Supporting Online Materials for

# Ultrastrong and Ultrastable Metallic Glass

Daisman P.B. Aji,[1] Akihiko Hirata,[1] Fan Zhu,[1] Liu Pan,[1] K. Madhav Reddy,[1] Shuangxi Song,[2] Yanhui Liu,[1] Takeshi Fujita,[1] Shinji Kohara[3] and Mingwei Chen[1,2,]*

[1] WPI-Advanced Institute for Materials Research (WPI-AIMR), Tohoku University
Sendai 980-8577, Japan
[2] State Key Laboratory of Metal Matrix Composites, School of Materials Science and Engineering, Shanghai Jiao Tong University, Shanghai 200030, PR China
[3] Japan Synchrotron Radiation Facility/SPring-8, Hyogo 679-5198, Japan

## Materials and Methods

**Preparation of bulk metallic glass.** The bulk metallic glass with a composition of $Zr_{55}Cu_{30}Ni_5Al_{10}$ (at. %) was prepared by copper-mold-casting. (*17*) Bulk specimens with a diameter of 50 mm were lathed from the cast. The DSC samples of 7 - 9 mg of bulk glass were obtained by cutting the bulk specimen with water cooled saw. The master alloy was prepared by arc-melting pure elements in a Ti-getter Ar atmosphere. The ingot was flipped and re-melted for several times to ensure chemical homogeneity, and then cooled slowly in the furnace.

**RF magnetron sputtering deposition.** The multicomponent glass films were fabricated by our recently-developed single-target rf magnetron sputtering. (*17, 18*) Ordinary and ultrastable metallic glass films were deposited directly on aluminum pans for DSC measurements. The sputtering target materials with a diameter of 50 mm and thickness of 3 mm were obtained from the bulk specimens with the composition of $Zr_{55}Cu_{30}Ni_5Al_{10}$ (at. %). To obtain free-standing samples, the films were deposited on (100) silicon substrates and then the silicon substrates were removed by dissolving them into a KOH solution. The base pressure of the deposition processes was lower than $10^{-4}$ Pa and argon was used as the ion source for sputtering. The deposition of film samples was performed by controlling the sputtering parameters: sputtering power, argon pressure, deposition temperature and cathode-to-sample distance. (*17*) The sputtering parameters were adjusted with power of 50 W, argon pressure of 0.3 Pa, and cathode-to-sample distance of 8 cm. The ordinary and ultrastable glasses were deposited at 300 and 573 K, respectively. With these sputtering parameters, the deposition rate obtained was 0.25 and 0.19 nm/s for ordinary and ultrastable glass, respectively.

**Differential scanning calorimetry (DSC) measurement.** A Perkin-Elmer DSC 8500 was used with argon as both a purge gas for the sample and a carrying gas for the intra-cooler. The instrument was calibrated with indium and zinc by using their melting points and their enthalpy of melting. During the course of measurements on a sample, the baseline, temperature calibration and stability of the equipment was frequently checked. For measurements of the samples with temperature higher than 993 K, high-temperature Extar DSC 6300 was used.

**X-rays Diffraction.** XRD profiles were obtained by Rigaku SmartLab x-rays diffractometer with Cu K-α radiation. High-energy XRD with the beam energy of 113.3 keV was carried out in transmission geometry at SPring-8 synchrotron radiation facility, Japan. The diffraction data were analyzed following the procedures in Ref. 27.

**Nanoindentation.** The hardness and elastic modulus of the bulk, ordinary and ultrastable glass were tested by MTS Nanoindentation$^{TM}$ G200 with a maximum load of 5 mN and a loading rate of 1 mN/s. The bulk sample was polished to make a mirror-like surface. For each sample, the hardness and elastic modulus were averaged from at least 25 measurements. The micropillars with nominal diameters ranging from ~2 μm to 4 μm and the aspect ratio of 2:1 (height: diameter) were prepared by a focused ion beam (FIB) system (JEOL JIB-4600F). (28) A 40 μm-in-diameter pool where the micropillar resides at the center was designed to provide a sufficient space for the indenter during a test. These micropillars are loaded in uniaxial compression at constant loading rate of 0.33 mN/s by using the ultra-micro-indentation system (Shimadzu W201S) with a 10 μm flat end Berkovoich indenter.

**TEM characterization.** Specimens of ultrastable glass for TEM characterization were prepared using sputtering deposition on holey carbon coated Cu grids. Structural characterization was performed using a JEM-2100F TEM (JEOL, 200 kV) equipped with double spherical aberration (Cs) correctors for both the probe-forming and image-forming lenses. HAADF images, in which the contrast is basically proportional to the square of the atomic number, were acquired using an annular-type STEM detector while BF-STEM images were simultaneously recorded using a STEM BF detector. The collecting angle ranges from 100 to 267 mrad, which is sufficient for HAADF-STEM. ABED patterns were recorded by a television-rate CCD camera (Gatan, UltraScan 1000). For ABED measurements, a nearly parallel coherent electron beam produced by a small condenser aperture with a diameter of 5 μm is used as a nanoprobe. The coherent electron beam can be accurately aligned and focused to a diameter as small as 0.4 nm. (20)

**Figure S1**

**Figure S1 ABED patterns of the ultrastable metallic glass.** (Beam diameter: 0.36 nm.) **a,** Twelve ABED patterns taken consecutively with a scan step of 1.2 Å along a line. One diffraction vector marked with an arrow on top side disappears at frame 9 and a new diffraction spot, marked with an arrow on the right side, appears at frame 6 and disappears at frame 12. **b and c,** ABED patterns of two regions shown in **a**. **b' and c',** Simulated ABED patterns for <0 2 8 1> polyhedra in $Zr_5Cu_3Al_3Ni$ glass. **b'' and c'',** Schematic atomic cluster of <0 2 8 1> polyhedra in $Zr_5Cu_3Al_3Ni$ glass. Based on the comparison between the experiment and simulation ABED patterns, as an example, the MRO domain in ultrastable glass can be indexed as a quaternary <0 2 8 1> cluster.

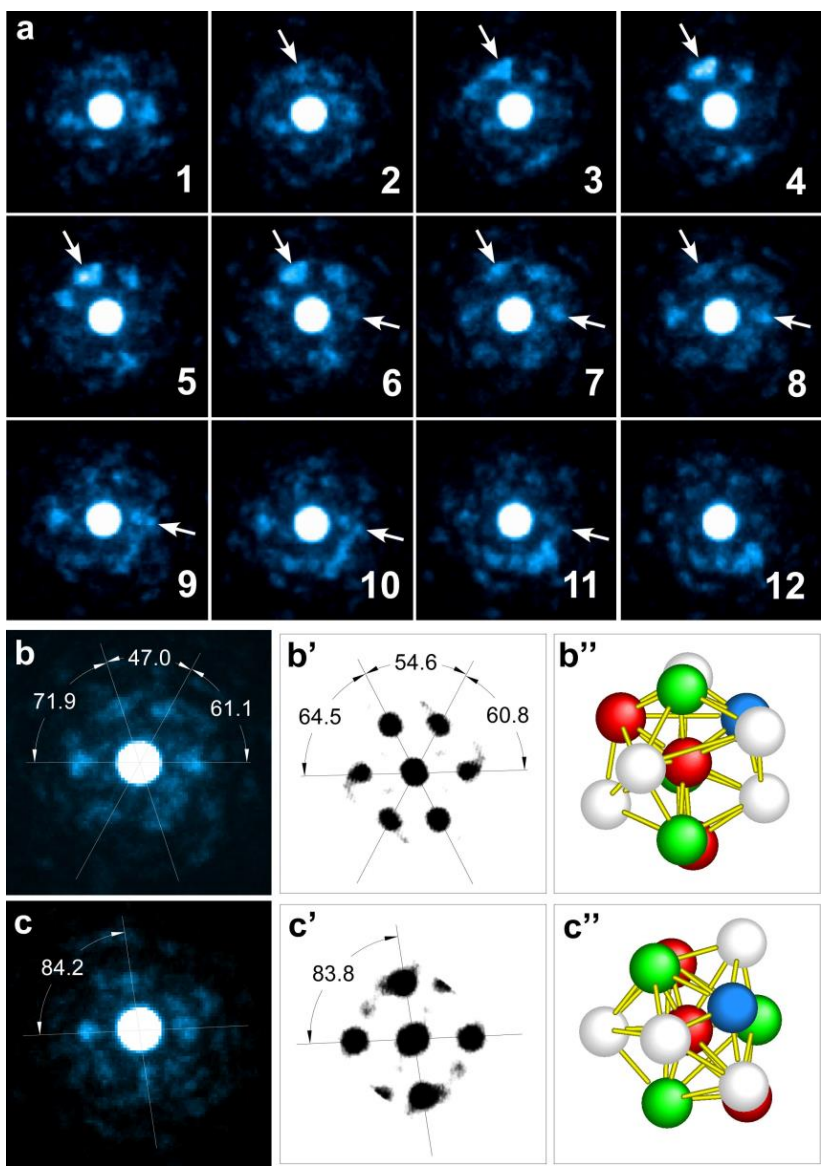

**Table S1**

**Table S1** The bond lengths ($r_{ij}$) between atoms and the calculated weight factors for the ZrCuNiAl glass.

| i-j | $r_{ij}$ (Å) | Weight factor | Heat of mixing (kJ/mol) |
|---|---|---|---|
| Zr-Zr | 3.16 | 0.377 | - |
| Zr-Cu | 2.85 | 0.345 | -23 |
| Zr-Ni | 2.86 | 0.080 | -49 |
| Zr-Al | 3.01 | 0.050 | -44 |
| Cu-Cu | 2.54 | 0.079 | - |
| Cu-Ni | 2.55 | 0.036 | -1 |
| Ni-Ni | 2.56 | 0.004 | - |
| Cu-Al | 2.70 | 0.023 | -1 |
| Ni-Al | 2.71 | 0.005 | -22 |
| Al-Al | 2.86 | 0.002 | - |

**All References**